\documentclass[11pt, letterpaper]{article}

\usepackage{graphicx}
\usepackage{amsmath}
\usepackage{amssymb}
\usepackage{float}
\usepackage{setspace}
\usepackage{epstopdf}
\usepackage{color}
\usepackage[letterpaper,left=1in,right=1in,top=1in,bottom=1in]{geometry}
\usepackage{multirow}
\usepackage{longtable}
\usepackage{rotating}
\usepackage{setspace}
\usepackage{geometry}
\usepackage{graphicx}

\usepackage{makeidx}
\usepackage{setspace}
\usepackage{epstopdf}
\usepackage{ragged2e}
\usepackage{caption}
\usepackage{hyperref}
\usepackage{mwe}
\usepackage{mathrsfs}
\usepackage{verbatim}
\usepackage{subfigure}
\usepackage{amsmath}

\usepackage{algorithm,algcompatible,amsmath}
\algnewcommand\INPUT{\item[\textbf{Input:}]}%
\algnewcommand\OUTPUT{\item[\textbf{Output:}]}%

\newcommand{\m}{\mathbb}

\title{\Large \bf Knowledge-based optimal irrigation scheduling of agro-hydrological systems}
\author{
\centerline{\normalsize Soumya Ranjan Sahoo$^{a}$, Bernard Agyeman, Sarupa Debnath, Jinfeng Liu$^{a,}$\thanks{Corresponding author: J. Liu. Tel: +1-780-492-1317. Fax: +1-780-492-2881. Email: jinfeng@ualberta.ca.}}
\vspace{5mm}\\
\centerline{\small $^{a}$Department of Chemical \& Materials Engineering, University of Alberta,} \\
\centerline{\small Edmonton, AB T6G 1H9, Canada}}
\allowdisplaybreaks

\begin{document}

\date{}
\maketitle


\begin{abstract}

The typical agricultural irrigation scheduler provides information on how much to irrigate and when to irrigate. The accurate and effective scheduler decision for a large agricultural field is still an open research problem. In this work, we address the high dimensionality of the agricultural field and propose a systematic approach to provide optimum irrigation amount and irrigation time for three-dimensional agro-hydrological systems. The water dynamics of the agro-hydrological system are represented using a cylindrical three-dimensional Richards Equation. We introduce a structure-preserving model reduction technique to decrease the dimension of the system model. Using the reduced model, the optimization-based closed-loop scheduler is designed in model predictive control (MPC) environment. The closed-loop approach can handle weather disturbances and provide improved yield and water conservation. The primary objective of the proposed scheduler is to ensure maximum yield, minimum water consumption and maximize the time between the two irrigation events, which results in less electricity usage. The proposed approach is applied to three different scenarios to show the effectiveness and superiority of the proposed framework. 
\end{abstract}

\newpage

\section{Introduction}
Freshwater scarcity is one of the most critical global risks the world is currently facing \cite{global_2015} due to mainly population growth, climate change and, environmental pollution. Almost 70\% of the total freshwater \cite{wastewater_2017} is consumed in agricultural irrigation every year. Moreover, the efficacy of the current irrigation methods is around 50\% to 60\% \cite{lozoya_model_2014} which is not adequate to save water usage significantly. Therefore, to alleviate water usage efficiency is a very important measure for feeding the growing population and managing the water crisis. 

The common irrigation practice includes open loop control which means no-real time information from the farm is used for the irrigation decision. Moreover, it is mostly depended on the farmers observation and experience about the farm instead of actual field condition such as soil moisture which may leads to excessive or insufficient irrigation. One of the solution to handle the issue is closed-loop irrigation where the real-time information from the field is utilized to make irrigation decision. There are different control strategies based on real-time field feedback have been mentioned in the literature. A modified
proportional-integral-derivative (PID) controller is designed for excellent regulation of soil moisture that responds to changes in environmental conditions quickly \cite{goodchild_method_2015}. Model predictive control (MPC) has gained increasing interest and attention in the literature in the past number of years as an optimal feedback control method. To optimize soil moisture and irrigation time, McCarthy et al. is thoroughly investigated a MPC using crop production models in \cite{mccarthy_simulation_2014}. Similarly, a MPC is designed by Delgoda et al. to minimize root zone soil moisture deficit and irrigation amount with a limit on water supply \cite{delgoda_irrigation_2016}. In the same line of research, a zone MPC algorithm is proposed to maintain soil moisture within a target zone instead
of a set-point control which shows further water conservation. These type irrigation control methods mainly emphasis on short period for couple of minutes or an hour of irrigation management. All these studies have one common objective to maintain the soil moisture deficit for a fixed set point or predetermined zone.

Irrigation scheduling is a well-defined planning for agriculture irrigation management to determine both the appropriate time and the quantity of water needs to irrigate fields while maximizing crop yields with the minimum water used. It is an architecture of water management over  over a much longer time period for e.g., the entire harvesting season of a crop. Typically, a water balance considers water consumption for plant growth, drainage and evaporation loss, past weather and climate information, soil moisture measurements of field in irrigation scheduling \cite{thorp_cotton_2017, j_cahoon_microcomputer-based_1990,jensen_scheduling_1970}. To maximise the expected profit, an optimal temporal allocation of irrigation water is designed using dynamic programming (DP) algorithm to determine optimal control policies at each irrigation decision point, conditional on the state of the system (soil moisture content) \cite{bras_intraseasonal_1981}. To maximize potential crop production through appropriate water allocation, quadratic programming (QP) optimisation is designed  based on real-time evaluations of crop water requirements \cite{wardlaw_optimal_1999}. For maximizing irrigation uniformity and minimizing yield reduction, a genetic algorithms (GA) is designed by Hassan et al. for optimal water allocation based on crop type, growing stage, and sensitivity to water stress using remotely sensed data \cite{hassan-esfahani_assessment_2015}. Irrespective of the objectives, in the aforementioned studies,  the scheduling optimization is only solved once at the beginning of the crop growing season, and the solution is used for the entire season.


In \cite{barnard_species-specific_2015}, the automated irrigation scheduling techniques was developed and implemented using the real-time prediction from MAESTRA (Multi-Array Evaporation Stand Tree Radiation A) model. The irrigation efficiency, tree growth and the amount of water supplied were compared with the model-based irrigation to the sensing-based irrigation. The author reported the growth of a tree in the model-based method is better than the sensing-based method. Recently, to optimize water allocation on both a daily and seasonal temporal scale, a hierarchical feedback approach is proposed by Nahar et al.to maximize yield over an entire growth season based on a linear parameter varying (LPV) model in \cite{nahar_closed-loop_2019}. A top-level scheduler evaluate set point in terms of required soil moisture for each remaining day in the growing season and a lower-level controller tries to follow that soil moisture set point. In \cite{kassing_optimal_2020}, Kassing et al. proposed a two-level optimization method that is used to solve an optimal irrigation problem of a large-scale plantation.  In this approach, the seasonal irrigation planner determines the optimal allocation of water over the fields for the entire growth season to maximize the crop yield whereas, a MPC takes care of the daily regulation of the soil moisture, considering the water distribution determined by the seasonal planner. To handle the nonlinearity of the system, \cite{bernard2022dycop} proposed a scheduling approach based on a data-driven model in the framework of MPC with discrete decision variables for one-dimensional agro-hydrological systems. One common feature of the above scheduling studies is that the scheduler and controller is designed using a simplified linear model which considers one-dimensional flow regime approach.

The three-dimensional model is necessary for precision irrigation because it can capture the runoff, horizontal flows in the presence of slopes and different soil types. An agrohydrological system can be modeled by three-dimensional Richards equation which is a partial differential equation (PDE) in nature. The finer discretization is required not only to ensure the numerical stability but also to ensure the local equilibrium between the soil water content and soil water potential \cite{babaeian_ground_2019}. The finer discretization produces very high number of nodes therefore, it increases the number of 
states of the discretized agricultural field as high as ($10^4-10^7$). The direct application of the state-space model (\ref{eq:nonlinear}) is computationally challenging for the optimization step in the MPC solving if states are the decision variables or the state constraints present. For example, there are three direct approaches to solve numerically the MPC optimization problem: single shooting, multiple shooting, and collocation. For nonlinear systems, multiple shooting and collocation-based approaches are more preferred because of the faster convergence and better numerical stability. However, the states of the system become the decision variables in both approaches. So for a large-scale system like an agro-hydrological system, the optimization problem may become intractable. Similarly, in Zone MPC, the presence of slack variables (twice the size of the states) as the decision variables, and the state constraints makes the optimization problem to a computationally dominant task.

Model reduction is one of the efficient techniques to handle the problem by reducing the dimension of states. Therefore, the number of decision variables and the number of state constraints decrease to a huge extent. A few popular classical model reduction techniques are proper orthogonal decomposition (POD), optimal Hankel norm reduction, balanced truncation methods and so on. \cite{antoulas_approximation_2005}. In the classical model reduction techniques, the state constraints can not be applied in the reduced-order dimension because the reduced states do not preserve any structure. There are some recent researches on the structure preserving model reduction where states preserve the structure \cite{ishizaki_model_2014,cheng_graph_2016,cheng_gramian-basedmodel_2019,sahoo_optimal_2019}. However, these methods are limited to only linear systems.

The primary objective of the work is to study the closed-loop scheduling in agricultural management. In this work, we propose an optimization-based scheduler design that provides the optimal time for irrigation events and optimum amount of irrigation. The objective of the scheduler is to maximize the crop yield while minimizing the water amount of irrigation.
First, the structure-preserving model reduction method is proposed based on system state trajectories of the agro-hydrological systems. Then the states which have similar trajectories are clustered based on unsupervised machine learning methods. The reduced-order model is generated based on the Galerkin projection using the clustering information. Based on the reduced-order model, the proposed closed-loop scheduling is built and investigated. In this framework, both irrigation time and irrigation water are considered as the decision variables. The effectiveness of the designed scheduler is thoroughly investigated using four different scenarios. 

The remainder of this work is organized as follows. Section 2 describes the water flow dynamics of agro-hydrological systems and crop models. Section 3 proposes the algorithms of structure-preserving model reduction methods. Section 4 
presents the closed-loop scheduler design.  In section 6, the proposed model reduction and scheduler designs are applied to an agricultural field under different scenarios. 

\section{System description and problem formulation}
\subsection{System description}

\begin{figure}[t]
\centering
\includegraphics[width=0.8\textwidth]{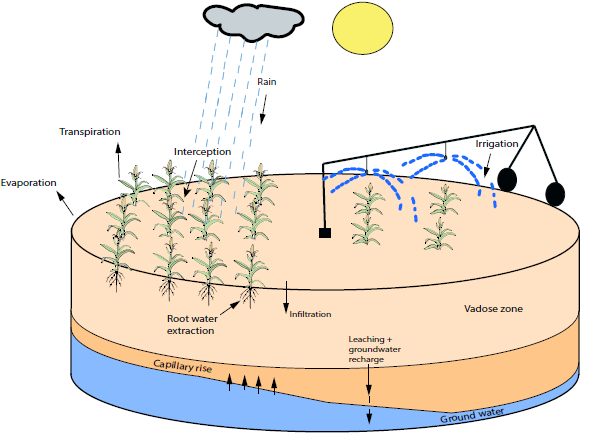}
\caption{Schematic of an agro-hydrological system \cite{agyeman_soil_2020}}
\label{fig:agro}
\end{figure}
We consider an agro-hydrological system that characterizes the interaction between the soil, the water, the atmosphere, and the crop. The schematic of the considered agro-hydrological system is shown in Figure~\ref{fig:agro}. The dynamics of the water flow in the agro-hydrological system can be represented by Richards Equation \cite{richards_capillary_1931} as follows: 
\begin{equation}
{\dfrac{\partial \theta}{\partial t}=c(h)\dfrac{\partial h}{\partial t}=\nabla \cdot(K(h)\nabla(h+z))+S(h,z)}
\label{eq:richards}
\end{equation}
where $h$ [m] is the field water pressure head, $\theta$ [m$^{3}$m$^{-3}$]  is the field water soil moisture content, $c(h)$ [m$^{-1}$] is the soil water capacity, $K(h)$ [ms$^{-1}$] is the hydraulic conductivity, $z$ [m] is the vertical coordinate, $S(h,z)$ [m$^{3}$m$^{-3}$s$^{-1}$] is the source and sink term consists of plant root water extraction. 

The details of the nonlinear relationship between hydraulic conductivity $K(h)$,  capillary capacity $c(h)$ and soil moisture content ($\theta$) with pressure ($h$) can be found in \cite{mualem_new_1976}, \cite{van_genuchten_closed-form_1980}, \cite{sahoo_optimal_2019}. 


The sink term $S(h,z)$ in (\ref{eq:richards}) characterizes the root water extraction rate. The total root water uptake depends upon transpiration rate, soil pressure head and root depth. The mathematical formulation of root-water uptake based on Feddes model is expressed as follows \cite{feddes1982simulation}: 
\begin{equation}
    S(h,z) = \alpha(h)S_{max}(z)  
\end{equation}
where $S_{max}(z)$[m$^{3}$m$^{-3}$s$^{-1}$] is the maximum possible water extraction rate under optimal condition. $\alpha(h)$[-] is the dimensionless water stress factor. The water stress factor depends upon the pressure head values and can be expressed as follows: 
\begin{equation}
  \alpha(h)=\left\{
  \begin{array}{@{}ll@{}}
    0, &  h>h_1 \\
    \frac{h1-h}{h1-h2}, & h_1> h\geq h_2 \\
    1, & h_2> h \geq h_3 \\
    \frac{h-h_4}{h_3-h_4}, & h_3 > h \geq h_4 \\
    0, & h_4> h
  \end{array}\right.
  \label{eq:alpha}
\end{equation} 
where $h_1$ is the pressure head value above which the root does not extract any water, $h_2$ and $h_3$ are the upper and lower threshold values between which the root water extraction is maximum. $h_4$ is the permanent wilting point, below which the root water extraction is zero. 
The maximum possible water extraction rate $S_{max}(z)$ can be calculated using Feddes model \cite{feddes1982simulation}: 
\begin{equation}
    S_{\text{max}}(h,z) = \frac{TP_p}{L}
\end{equation}
where $TP_p$ [ms$^{-1}$] is the potential transpiration rate and $L$ [m] is the rooting depth. The potential transpiration rate $TP_p$ is computed by:  
\begin{equation}
    T_p = ET_p- EV
\end{equation}
where $ET_p$ [ms$^{-1}$] is the potential evapotranspiration rate and $EV$ [ms$^{-1}$] is the potential evaporation rate. The $EV$ can be expressed as follows: 
\begin{equation}
   EV = ET_p*e^{(-0.623LAI)}
\end{equation}
where $LAI$ is the leaf area index. The potential evapotranspiration 
$ET_p$ is computed as follows:
\begin{equation}
    ET_p = K_cPET
\label{eq:ET}
\end{equation}
where $PET$ [ms$^{-1}]$ is the reference evaporation rate which can be calculated based on the Penmon-Moneith equation \cite{allen_crop_nodate} and $K_c$ [-] is the crop coefficient. 

The crop yield model is also important along with the Richards equation. The crop yield model is modeled as a function of actual and potential evaporation and as follows: 
\begin{equation}
 \bigg(1-\frac{Y_a}{Y_p}\bigg) = \sum_{k=1}^T K_y(k) \bigg(1-\frac{ET_a}{ET_p}\bigg)
 \label{eq:crop}
\end{equation}
where $Y_a$ is actual yield and $Y_p$ is potential yield. $K_y(k)$ is the crop sensitivity factor at time $k$. $T$ is the total time for growing seasons. $ET_a$ is the actual evapotranspiration. $ET_a$ can be represented as: 
\begin{equation}
    ET_a = \alpha(h)ET_p
    \label{eq:ETa}
\end{equation}
where $\alpha(h)$ is calculated from Equation \ref{eq:alpha} and $ET_p$ is calculated from Equation \ref{eq:ET}.  

In Equation \ref{eq:crop}, $(1-\frac{Y_a}{Y_p})$ represents the yield deficiency and  $(1-\frac{ET_a}{ET_p})$ represents the $ET$ deficiency. By combining the Equation \ref{eq:ETa} and Equation \ref{eq:crop}, the relationship between the crop yield and soil pressure head is expressed as follows: 
\begin{equation}
     \bigg(1-\frac{Y_a}{Y_p}\bigg) = \sum_{k=1}^T K_y(k) \bigg(1-\alpha(h)\bigg)
 \label{eq:crop_a}
\end{equation}

In this work, we consider that the agricultural field is equipped with a center pivot irrigation system. A center pivot irrigation system rotates across the field around a fixed pivot at the center of the field and irrigates in a circular manner. In order to account for the circular movement of the center pivot irrigation system, the Richards equation in (\ref{eq:richards}) is expressed in the cylindrical coordinates as follows \cite{agyeman_soil_2020}: 

\begin{equation}
    c(h)\dfrac{\partial h}{\partial t}= \frac{1}{r}\frac{\partial}{\partial r}\Bigg[rK(h)\frac{\partial h}{\partial r}\Bigg]+
    \frac{1}{r}\frac{\partial}{\partial \theta}\Bigg[\frac{K(h)}{r}\frac{\partial h}{\partial \theta}\Bigg]+
    \frac{\partial}{\partial z}\Bigg[K(h)\Bigg(\frac{\partial h}{\partial z}+1\Bigg)\Bigg]+S(h,z)
\label{eq:cyl_richards}
\end{equation}

The three-dimensional agro-hydrological model in (\ref{eq:cyl_richards}) is a nonlinear partial differential equation, which renders the problem difficult to solve analytically. In this work, we apply the explicit finite difference method to discretize the Richards equation in (\ref{eq:cyl_richards}). Note that spatial discretization of the model is performed, such that a continuous-time state-space model is established as in the following form:
\begin{equation}
\begin{aligned}
\dot{x}(t)&= f(x(t), u(t))
\end{aligned}
\label{eq:nonlinear}
\end{equation}
where $x(t)\in \mathbb{R}^{N_x}$ denotes the states vector  representing the pressure head value at each discretized node of total size $N_x$ and $u \in \mathbb{R}^{N_u}$ represents the input vector containing $N_u$ irrigation values applied at each surface discretized node. As the input (irrigation amount) is applied to each surface node, it is incorporated in the system surface boundary condition. The surface boundary condition is characterized by Neumann boundary condition as follows:
\[
\frac{\partial{h}}{\partial{z}}\Bigg|_{r,\theta,z= Z_s} = -1 - \frac{u(t)}{K(h)}
\]
where $h$ is the pressure head, $u(t)$ is the input to the system, $K(h)$ is the hydraulic conductivity and $Z_s$ is the length of the soil column. The bottom boundary condition is specified as free drainage.

\subsection{Problem formulation}
As described in the previous section, the input $u$ is only present at the top surface nodes. We also discussed that we consider a real agricultural farm which is facilitated with the central pivot irrigation system. The circular motion of the central pivot allows to cover the entire farm area in a specified time i.e, the central pivot is not able to provide water to every plant at the same time. There is a rotational time gap for plant in receiving water. Therefore, we can say that that, at any instance, aligned to the central pivot where it irrigates, the current receiving region has some inputs $u$. As the whole region is discretized into a set of nodes, the current receiving region has nonzero nodes. Otherwise, the rest of the surface nodes are zero which do not have any input. At each time, only some nodes has inputs and it is changing over the rotation of the central pivot. Thus, this imposes a time-varying constraint on $u$ as follows: 
\begin{equation}
    \mathscr{U}_{lb}(t) \leq u(t) \leq \mathscr{U}_{ub}(t)
\end{equation}

A continuous time state-space model with measurements and disturbances is considered as follows: 
\begin{equation}
\begin{aligned}
\dot{x}(t)&= f(x(t), u(t), d(t)) \\ 
y(t)&  = Cx(t)
\end{aligned}
\label{eq:statespace}
\end{equation}
where $y(t)\in \mathbb{R}^{N_y}$ denotes the soil pressure head measurements, $d(t)\in \mathbb{R}^{N_d}$ is the weather disturbances.  

The final objective is to calculate the optimum time and irrigation amount for maximum crop yield and water conservation for a three-dimensional field model with a central pivot irrigation system. 

\section{Proposed model reduction}
\begin{figure}
\centerline{\includegraphics[width=0.8\columnwidth]{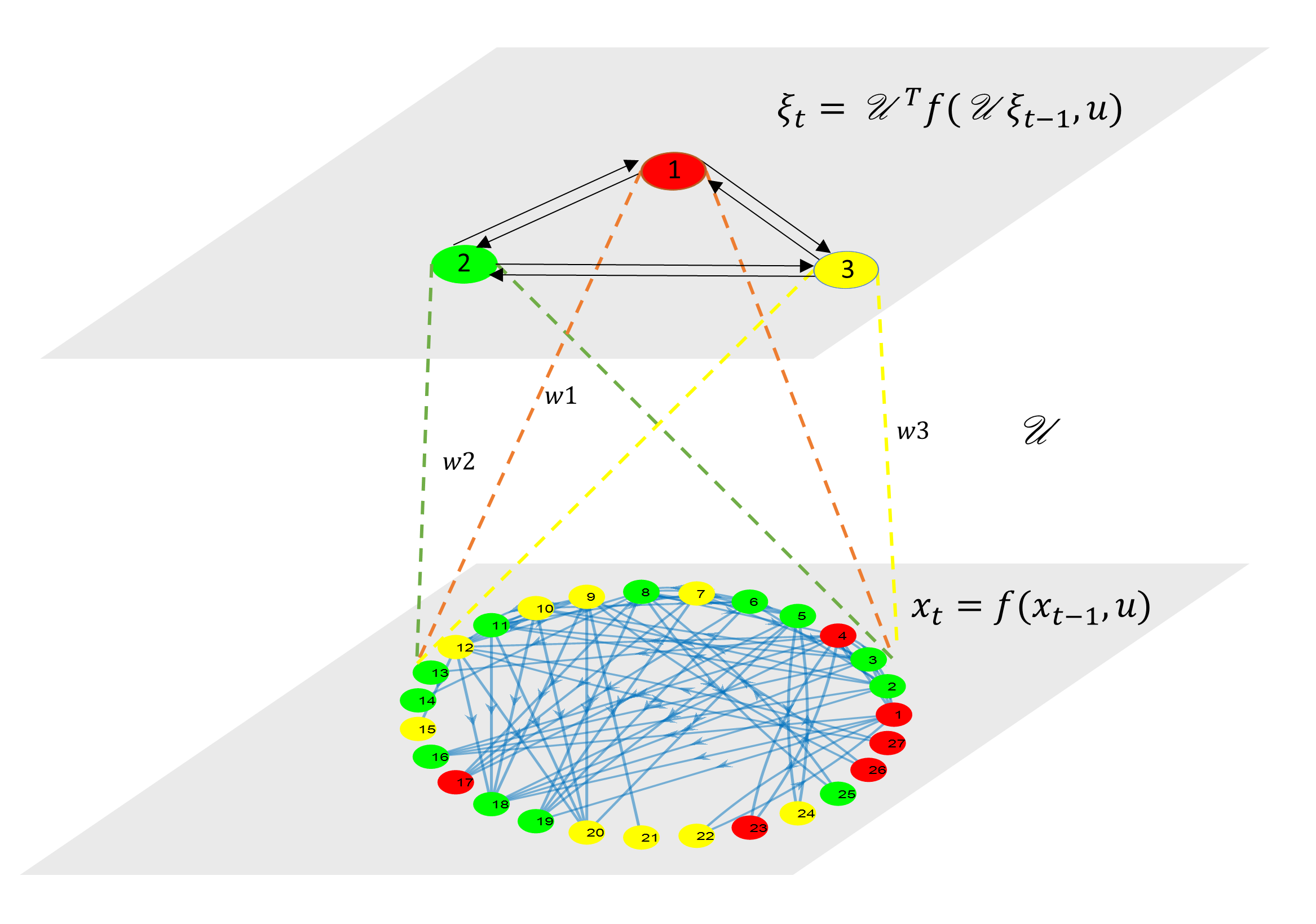}}
\caption{Illustration of structure-preserving model reduction.} 
\label{fig:mr}
\end{figure}

As discussed in the introduction, the finer discretization results in a large number of states for a large system. Thus the use of the state-space model (\ref{eq:nonlinear}) is computationally expensive for the optimization step in the scheduler design where states are the decision variables in multiple shooting and collocation based methods. Further, the optimisation cost increases if the necessary and important state constraints are added for the system. The model reduction can deal with these issues. In classical model reduction techniques, the states do not preserve the structure, so applying state constraints in a reduced model is difficult. Hence, we propose structure preserving model reduction technique. 
The calculation of a linear model in a large-scale system is also computationally expensive. So in this work, we propose the trajectory-based model reduction techniques. The first step is to obtain the system trajectories. The second step includes the cluster calculations based on similarity among the trajectories. The third step calculates the projection matrix and constructs the reduced-order model using the Petrov-Galerkin projection. The illustration of the proposed model reduction is shown in Figure \ref{fig:mr}. The details of each step in the proposed algorithm are described as follows. 

\subsection{Step 1: Snapshot matrix generation}
 The first step is to generate the state snapshots. Based on the prescribed input from the scheduler, simulate the nonlinear system (\ref{eq:nonlinear}) and generate the state trajectories from the initial time to the final time as follows: 

    \[\mathscr{X} =[x(t_{0}) ~x(t_{1})~\dots ~x(t_N)]\]
where $\mathscr{X}$ $\in \mathbb{R}^{n\times N}$  is the snapshot matrix of the actual system, $n$ is the number of states, $N$ is the total number of the sampling intervals. 
   
\subsection{Step 2: calculation of cluster sets}
In this step, the cluster matrix sets are generated. The main idea is to create clusters of states having similar dynamics based on the system trajectories. Then the projection matrix is generated using the clustering information and using the projection matrix, the system is projected from a higher dimensional system to a lower-dimensional reduced system. In this work, agglomerative hierarchical clustering \cite{steinbach_comparison_2000} is used. We use the Euclidean distance between trajectories as the distance measure for states. The main reason to choose agglomerative hierarchical clustering is because of the capability to define the distance threshold between the clusters instead of predefining the number of cluster sets. The distance threshold is a tuning parameter for the accuracy of the reduced model. There are three commonly used linkage methods present in agglomerative hierarchical clustering (e.g. single, average, complete linkage). In this work, we use the average linkage, and it considers the average distance between each point in one cluster to every point in other clusters. 
\[
d(p,q) = \frac{1}{n_pn_q}\sum_{i = 1}^{n_p}\sum_{j = 1}^{n_q}d(x_{pi},x_{qj}) 
\]
where $p$ and $q$ are two clusters, $i$ and $j$ are data points within the clusters, $d$ is the euclidean distance between $i$ and $j$ and $n_p,n_q$ are the size of the clusters of $p$ and $q$ respectively. 

Let us consider $\mathscr{C} = \{\mathscr{C}_1,\mathscr{C}_2, \dots, \mathscr{C}_r\}$ be the collection of clusters after the hierarchical clustering and $r$ is the order of the resulted reduced model. The resulted clusters have following properties: i) $\mathscr{C}_i \cap \mathscr{C}_j = \Phi$ and ii) $\mathscr{C}_1 \cup \mathscr{C}_2 \cup\ldots \cup \mathscr{C}_r =N_x$, where $N_x$ is the total number of states. 

\subsection{Step 3: reduced model construction}
In this step, reduced-order system is constructed based on the Petrov-Galerkin projection framework \cite{antoulas_approximation_2005}. For the Petrov-Galerkin projection method, the projection matrix is required. After the construction of state clusters, the projection matrix $\mathscr{U}$ is generated. 
The projection matrix is defined as
$\mathscr{U} \in \mathbb{R}^{n\times r}$, whose elements are expressed as follows:
\[\begin{aligned}
\mathscr{U}_{i,j}=\left\{
\begin{array}{@{}ll@{}}
w_i, & \text{if}\ \text{vertex} \ i \in \mathscr{C}_j \\
0, & \text{otherwise}
\end{array}\right.
\end{aligned}
\]
and $\mathit{w_i}$ is determined as follows:
\[
\begin{aligned}
\mathit{w_{i}}&= 1/||\alpha_i|| \\
\alpha_i&= \mathbb{E}_i ^{T}\alpha
\end{aligned}
\]
where $\alpha = [1,\dots,1]^T \in \mathbb{R}^{n}$, $||\alpha_i||$ is the $L_2$ norm of $\alpha_i$, $\mathbb{E}_i = e_{\mathscr{C}_i}\in \mathbb{R}^{n\times m}$, $e_j$ is the $j^{th}$ column of the identity matrix of size $\mathbb{R}^{n\times n}$ and $m$ is the cardinality of $\mathscr{C}_i$ set.

 The adaptive reduced model of (\ref{eq:nonlinear}) is expressed as:
\begin{equation}
\begin{aligned}
\dot{\xi}(t)&= f_r(\xi(t), u(t)) 
\end{aligned}
\label{eq:red_nonlinear}
\end{equation}
where $f_{r}(\xi(t), u(t))= \mathscr{U}^Tf(\mathscr{U}\xi(t), u(t))$ and $\xi(t) = \mathscr{U}^Tx(t)$. Note that the actual state $x(t)$ can be approximated based on mapping $\tilde{x}(t) = \mathscr{U}\xi$. The discrete model of (\ref{eq:red_nonlinear}) is expressed as follows: 
\begin{equation}
\begin{aligned}
\xi(k+1)&= f_{rd}(\xi(k), u(k)) 
\end{aligned}
\label{eq:red_nonlineard}
\end{equation}

The proposed reduced order model can be used in different applications. The main advantages of using the reduced-order system are as follows:
\begin{enumerate}
 \item Proposed method preserves the connection structure among subsystems while the general model techniques (POD, auto-encoder) does not preserve the structure
    \item Due to preservation of structure, the proposed model reduction can be used in the distributed state estimation, distributed controller design and sensor placement
    \item The actual state constraints can be used for the reduced model using the projection matrix 
    \item The proposed model reduction technique decreases the number of variables for a large scale system which makes the optimization problem tractable
    \begin{itemize}
        \item Zone MPC - a reduction in the number of slack variables and constraints
        \item Collocation based method or multiple-shooting methods - a reduction in the number of state variables
    \end{itemize}
    \item The calculation of the jacobian matrix (which is useful for optimization and linearization) for the system is reduced to a very great extent. 
\end{enumerate}

\section{Proposed closed-loop scheduling}
\label{section:estimation}
This section proposes closed-loop scheduling to calculate the time intervals between irrigation events and the water amount for each event. The primary objective is to maximize the crop yield while reducing the total water use and equipment operating cost. The scheduler considers historical weather and weekly weather forecast, and soil moisture measurements. The main idea is to irrigate the field and calculate the time required to reach the lower stress-free zone. 

In this work, the iterative finite horizon optimization is considered like the classical MPC. However, the length of the horizon is not fixed because, the time is also a decision variable in the optimization problem. The maximum time can be added in the constraint as higher bound. The optimizer can make a prediction till the higher bound of time period and decide what is the optimum time within that limit. However, the weather forecast till the higher bound of time may not be accurate enough. So the receding horizon strategy is implemented to handle the uncertainty in the weather forecast. The optimization problem is resolved after few days interval with a more accurate weather forecast and recent measurements. 

Each horizon consists of three separate segments. In the first segment, we irrigate the field, and the amount of water to be irrigated is the primary decision variable. In the second segment, the time is a decision variable that calculates the time for the next irrigation event. The time is calculated such that the plants will not experience stress, and the yield will be maximized. The third segment calculates the irrigation amount for the next horizon. The third segment is added to the optimization problem to give the optimizer some flexibility to see few more days future forecasts and make the irrigation time and amount more robust. In all three segments, the yield and the constraints are considered to keep the pressure head within a stress-free zone. The slack variables are introduced to relax the target zone.

\begin{figure}[!t]
\centerline{\includegraphics[width=0.7\columnwidth]{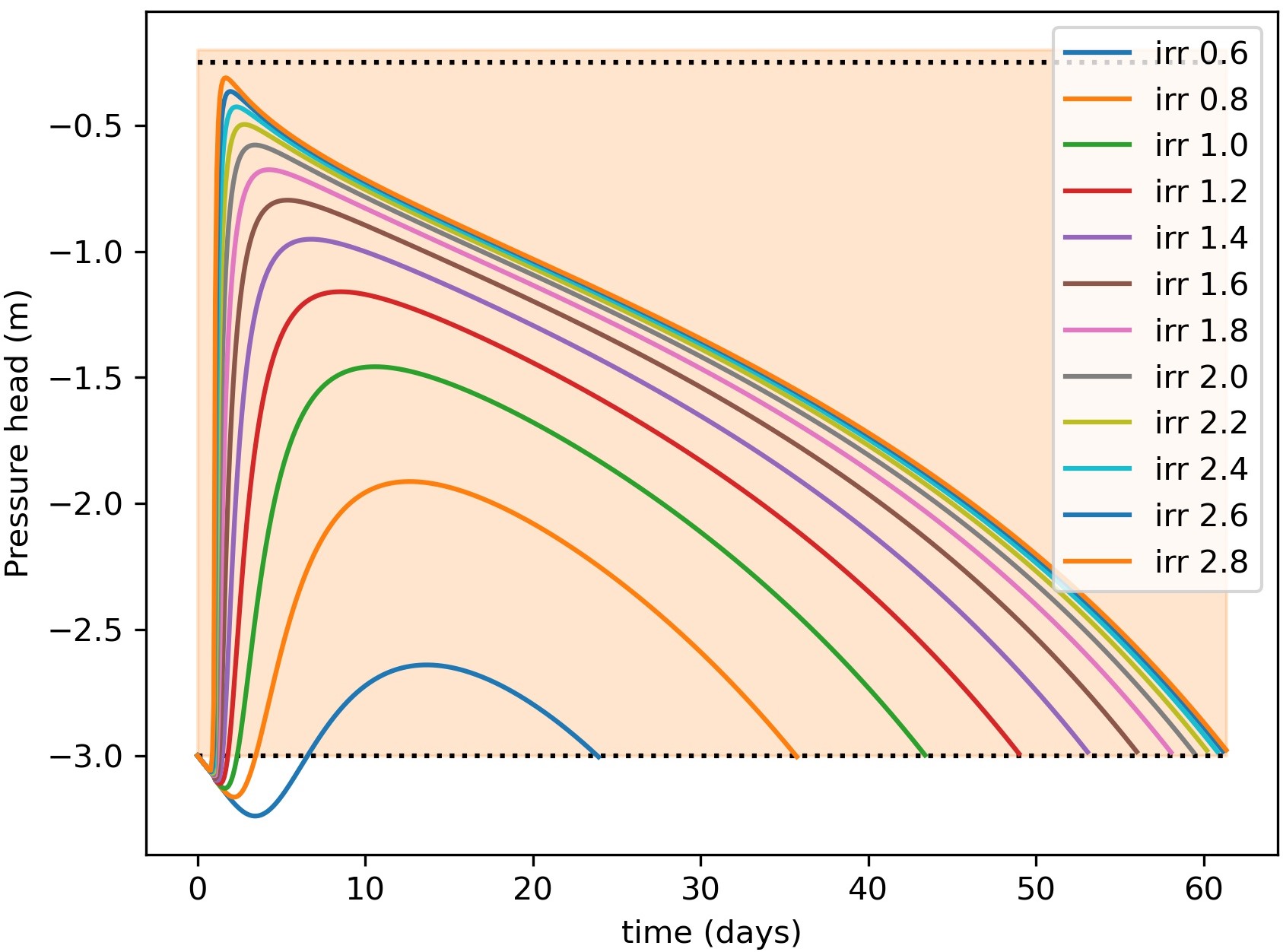}}
\caption{Motivation of optimizing both input and time together} 
\label{fig:motivation}
\end{figure}

All three segments are optimized simultaneously. The reason to optimize everything together is that the amount of irrigation in the first and third segments will affect the total time in the second segment. The scenario may arise where we irrigate enough, but the total time for the next irrigation does not change much. This can be illustrated using a simple example. Let us consider the central pivot irrigation system. The irrigation amount is changed gradually from $0.6\times10^{-6}$ m/s to  $2.8\times10^{-6}$ m/s as shown in Figure \ref{fig:motivation}. First, we irrigate the system and calculate the time when it reaches the lower zone. From Figure \ref{fig:motivation}, we can observe that with irrigation amount $0.6\times10^{-6}$ m/s, the number of days it reaches the lower zone in 23 days, and with irrigation amount $2.8\times10^{-6}$ m/s, it takes around 61 days. It can also be observed that after $1.6\times10^{-6}$ m/s irrigation amount, the number of days to reach the lower zone does not change much. Even if putting double the amount of water, the next irrigation days do not change significantly. So if we optimize both the irrigation amount and the time together, we can save water and optimize the equipment functioning cost. 


For each horizon, the optimization problem is formulated as follows: 

\begin{subequations} 
\begin{align}
\min\limits_{u(j),\overline{\epsilon_r}(j), \underbar{$\epsilon_r$}(j), T} &  Q_y\Bigg(1-\frac{Y_a}{Y_p}\Bigg)^2+ Q_u\sum\limits_{j=k}^{N-1}u(j) - Q_T\sum\limits_{j=k+N_1}^{\tiny{N_1+N_2}}\frac{T}{T_{ub}} + \sum\limits_{j=k+1}^{k+N}(\overline{Q_r}\overline{\epsilon_r}(j)^2+\underbar{$Q_r$}~ \underbar{$\epsilon_r$}(j)^2)\vspace{2mm} \label{eq:mpc_a}\\
{\rm s.t.~} &\Bigg(1-\frac{Y_a}{Y_p}\Bigg) = \sum\limits_{j=k+1}^{k+N} K_y(j)(1-K_s(y_r)) \label{eq:mpc_b} \\
&\tilde \xi(j+1) = \mathscr{U}^Tf(\mathscr{U}\tilde \xi(j),u(j),d),~~ j = k,...,N-1 \vspace{2mm} \label{eq:mpc_c}\\
& y_r(j) = C_r\tilde \xi(j) \vspace{2mm} \label{eq:mpc_d}\\
& u_{lb} < u(j) < u_{ub}, ~j = k,...,k+N_1 ~\& ~k+N_1+N_2,....k+N_1+N_2+N_3 \vspace{2mm}\label{eq:mpc_e}\\ 
& u(j) = 0, ~~j= k+N1,...,k+N1+N2 \vspace{2mm} \label{eq:mpc_f}\\
& \tilde \xi(j) \in \m Z, ~~j= k,...,k+N-1 \vspace{2mm} \label{eq:mpc_g}\\
& \underbar{$V$} - \underbar{$\epsilon_r$}(j) < y_r(j) < \bar{V} + \bar{\epsilon_r}(j) \vspace{2mm} \label{eq:mpc_h}\\
&  \bar{\epsilon_r}(j) \geq 0, \underbar{$\epsilon_r$}(j) \geq 0 \vspace{2mm} \label{eq:mpc_i}\\
& T_{lb} < T < T_{ub} \vspace{2mm} \label{eq:mpc_j}\\
& N_1+N_2+N_3 = N\vspace{2mm} \label{eq:mpc_k}
\end{align}
\label{eq:mpc}%
\end{subequations}
where Equation (\ref{eq:mpc_a}) defines the cost function to be minimized and the input ($u$), time ($T$) and slack variables ($\overline{\epsilon_r}, \underline{\epsilon_r}$) are the decision variables. In Equation (\ref{eq:mpc_a}), the first term is the crop yield deficiency cost, the second term considers the irrigation water cost, the third term denotes the normalized time cost which is active only in second segment. The last term in Equation (\ref{eq:mpc_a}) is the cost term of non-negative slack variables ($\overline{\epsilon_r}, \underline{\epsilon_r}$) which is introduced to relax the bounds of target zones $\bar{V}, \underline{V}$ in Equation (\ref{eq:mpc_h}). 
$Q_y, Q_u, Q_T, \underline{Q_r}, \overline{Q_r}$ are the positive weighting factors.
Equation (\ref{eq:mpc_b}) is the model used to evaluate the yield deficiency. Equations (\ref{eq:mpc_c},\ref{eq:mpc_d}) represent the discrete time reduced-order model and the output function. $\tilde \xi(k)$ denotes current state estimates at time $k$. In this work, we assume all the states can be estimated. 
 $N_1$ is the number of sampling time for first segment, $N_2$ is the sampling time for second segment ($\Delta T_2 = T/N_2$). Things to note that in segment 2, the time is unknown so the number of sampling points ($N_2$) is fixed and it is chosen based on upper bound of $T_{ub}$ such that the model does not experience numerical issues. $N_3$ is the number of sampling time for third segment. Equation (\ref{eq:mpc_k}) shows the total sampling time is $N$. Equation (\ref{eq:mpc_e}) provides the bounds of input for first segment and third segment. Equation (\ref{eq:mpc_f}) defines the input amount is zero for second segment. Equation (\ref{eq:mpc_h}) imposes the zone constraints with the slack variables and Equation (\ref{eq:mpc_i}) implies the slack variables are non-negative.  Equation (\ref{eq:mpc_j}) defines the constraints for the lower and upper bound of time. 

As discussed before, the receding horizon strategy is implemented to handle the weather uncertainty and use the scheduler as a closed-loop system. The day ($T_s$) till which we can predict the accurate weather prediction is selected. In general we predict the weather for 7 days. So if the scheduler predict the next irrigation event to more than 7 days then, we reevaluate the scheduler optimization after 7 days again with the current field condition as initial condition and future 7 days weather prediction. If the scheduler predict the irrigation event less than 7 days, then we solve the optimization problem for next horizon using the recent day field condition as initial condition. The algorithms for the receding horizon is as follows: 
\begin{enumerate}
    \item At the current time ($k$) solve the optimization problem (\ref{eq:mpc}), with initial condition $\xi(k)$ and obtain the optimum input ($u$) and time ($T$). 
    \item If the optimum time ($T$) is greater than $T_s$ ($T>T_s$) then resolve the optimization problem (\ref{eq:mpc}) with initial condition $\xi(k+T_s)$ and obtain the optimum input and time. 
    \item Else ($T<T_s$) solve the optimization problem (\ref{eq:mpc}) with initial condition $\xi(k+T)$ and obtain the input and time. 
\end{enumerate}

\section{Results}
In this section, the proposed algorithms are applied to demonstrate the performance of reduced order model and the scheduler under different scenarios. A field of radius 50 m and depth 30 cm is considered. The field is equipped with central pivot irrigation system. The model of the farm is constructed using finite difference discretization of the Richards equation. The entire system is discretized into 1920 nodes with 5 in radial, 64 in azimuthal and 6 in axial direction. Each nodes corresponds to states of the system. The central pivot takes around 8 hours to irrigate the whole field. Different types of crops and soil types are considered in different scenarios. 

\subsection{Results:  model reduction} \label{sec:model_red}
In this subsection, the proposed model reduction discussed in section 3 is applied to the system. First, the effect of reduced model order on the mean square error (MSE) of the reduced model is discussed. Then the robustness of the reduced model is discussed. 

\begin{figure}[!ht]
\centering
\includegraphics[width=0.5\columnwidth]{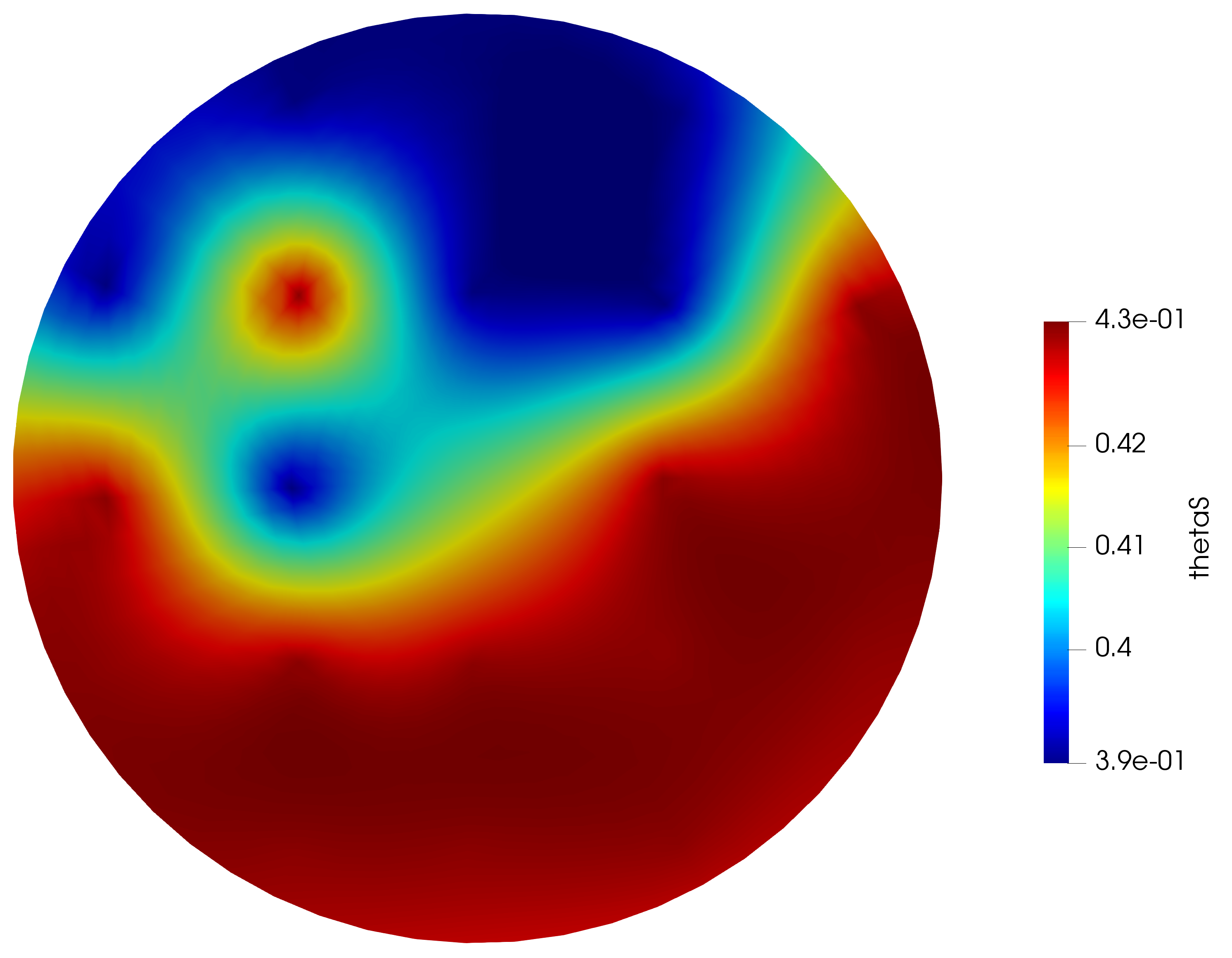}
\caption{Soil parameter $\theta_s$ for the field}
\label{fig:para}
\end{figure}

In this simulation, we consider the real soil properties of the field located at Lethbridge, Canada. In summer 2019, we collected soil samples at 20 points in the field and in the soil lab the soil types are estimated. We found three different soil types present in the field: loam, sandy clay loam, and clay loam. The hydraulic properties of the soil types are shown in Table \ref{tbl:soil_prop}. The kriging method is applied to get the soil properties all other nodes of the field. Figure \ref{fig:para} shows one selected parameter ($\theta_s$) of the surface nodes. The other parameters also follow the same trend. 

\begin{table}[t]
	\caption{Soil properties of three different types of soil}
	\small 
	\centering
	\begin{tabular}{|c|c|c|c|c|c|c|}
		\hline
		  Soil type & $K_{s}$ (m/s) &{$\theta_{s}$ $(\text{m}^{3}/\text{m}^{3})$}&{$\theta_{r}$ $(\text{m}^{3}/\text{m}^{3})$}& {$\alpha$ (1/m)}& {$n$ (-)}\\
		\hline
		 Loam & $2.889\times 10^{-6}$  & 0.43 & 0.0780&3.6& 1.56\\ 
		\hline
		  Sandy clay loam & $3.6388\times 10^{-6}$  & 0.39 & 0.1 &5.9& 1.48\\ 
		 \hline
		  Clay loam & $7.2223\times 10^{-7}$  & 0.41 & 0.095 &1.9 & 1.31\\ 
		  \hline
	\end{tabular} \label{tbl:soil_prop}
\end{table}

\begin{figure}[!ht]
\centering
\includegraphics[width=1\columnwidth]{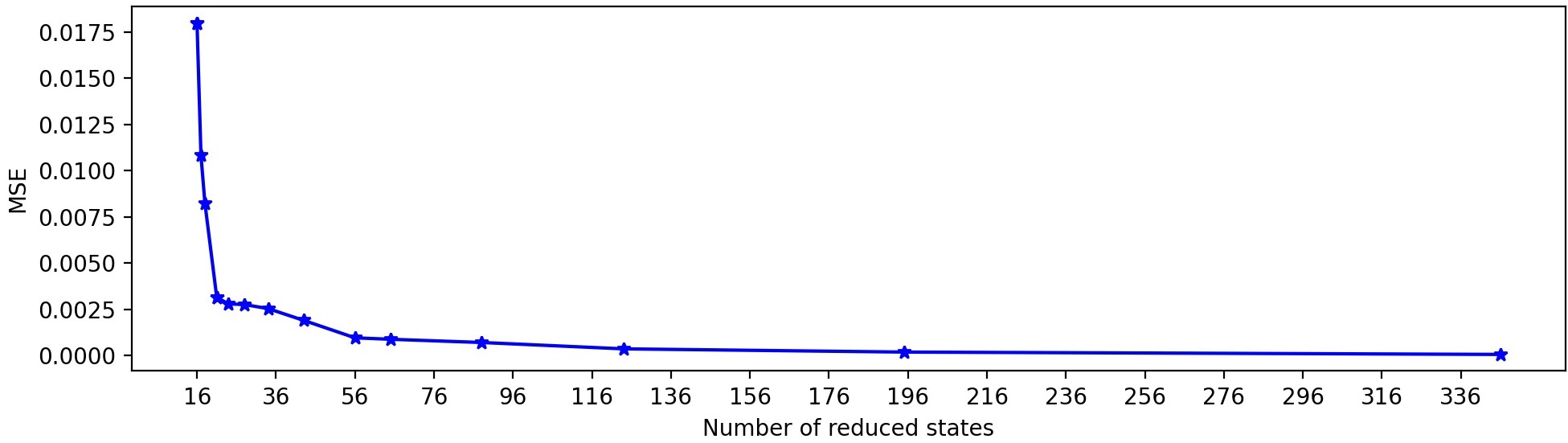}
\caption{Values of MSE with different reduced order}
\label{fig:Nc_case0}
\end{figure}

Figure \ref{fig:Nc_case0} shows the MSE values of the reduced model with the number of reduced states. The number of reduced states is obtained by changing the threshold values of the agglomerative hierarchical clustering method starting from 0.3 to 3.5 by increasing 0.2. As discussed in section 2, we can specify the threshold values in hierarchical clustering instead of the number of reduced models. The MSE values are calculated between the actual model and the reduced-order model. From Figure \ref{fig:Nc_case0} it can be observed that after 56 reduced states, the error values are minimal. 
For this simulation, we consider 30 reduced states. 

\begin{figure}[!ht]
\centering
\includegraphics[width=1\columnwidth]{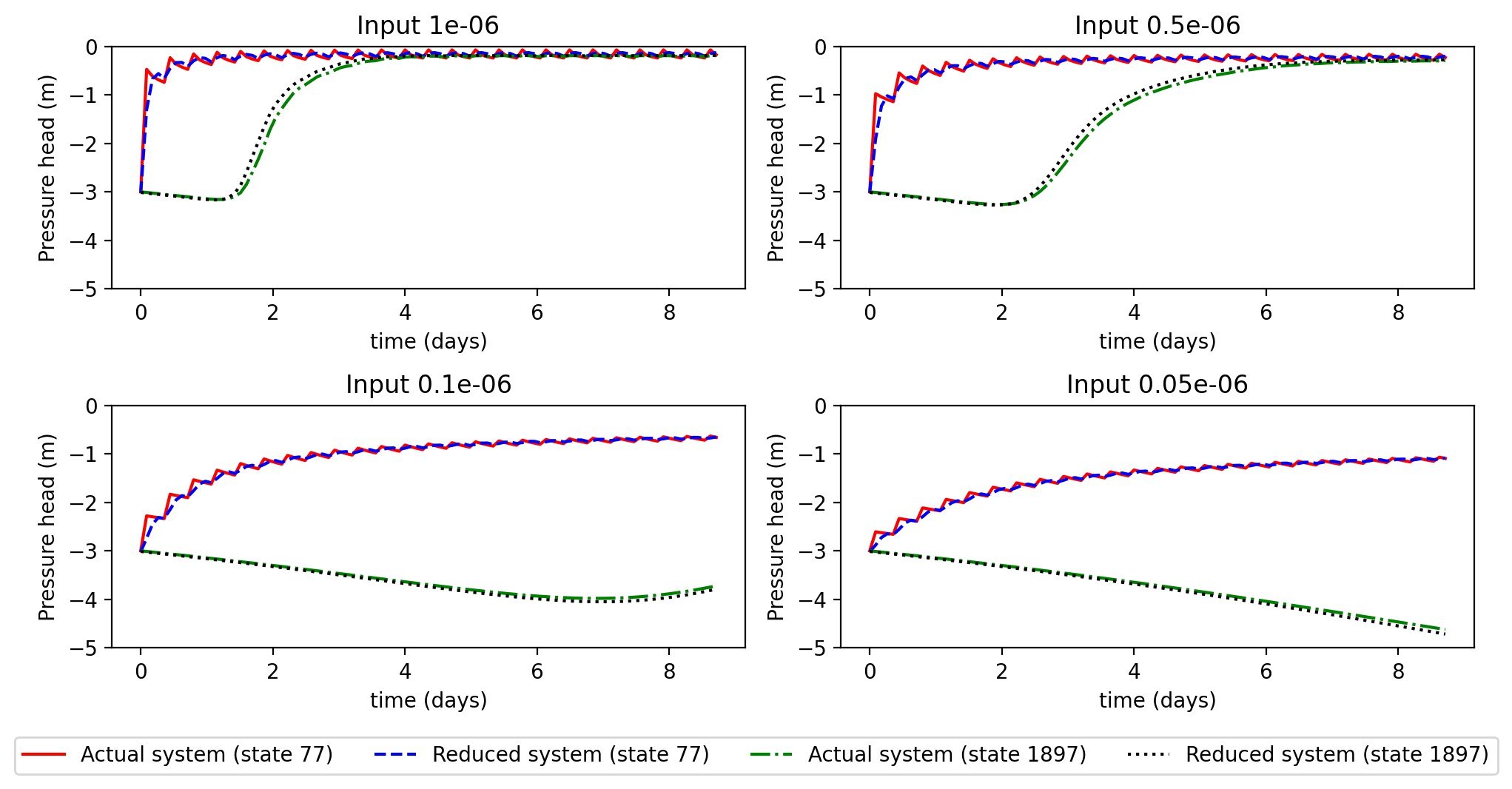}
\caption{Selected state trajectories of the actual system and reduced system for four different inputs}
\label{fig:compare_traj_case0}
\end{figure}

Next, we have presented the robustness of the reduced model. In optimization-based controller design, the input amount is one of the important decision variables. So the reduced model should be robust enough to handle different input amounts. First, the projection matrix is calculated using the initial condition -4.0 m and input amount 2e-06 m/s. Then using the same projection matrix, we simulate the reduced-order model starting from different initial condition (-3.0 m) and different input amounts (1e-06 m/s, 0.5e-06 m/s, 0.1e-06 m/s, 0.05e-06 m/s). Figure \ref{fig:compare_traj_case0} shows the state trajectories of the actual system and the reduced-order system randomly chosen state 77 (surface) and 1897 (depth 25 cm) for different input amounts. It can be observed that reduced order trajectories are very close to the actual order trajectories, which shows the robustness of the proposed model reduction method.

\subsection{Result: scheduler}
In this subsection, the performance of the proposed scheduler design is demonstrated under the following three scenarios: 
(1) Scenario 1: uniform soil types, no rain, constant ET and grass; 
(2) Scenario 2: non-uniform soil types, no rain, constant ET and grass; (3) Scenario 3: uniform soil, dry bean crop type, variable ET and rain. 
\subsubsection{Scenario 1}
In this subsection, we consider a simple scenario of uniform soil type of loam. We further assumed that there are no disturbances present like rain and variable ET. We consider the crop as grass. The simple scenario is considered to show how the proposed scheduler algorithm works. 

We consider the $2^{nd}$ layer from the top as the root zone of the grass, which is the output to the optimization problem (Equation (\ref{eq:mpc_d})). The values of the tuning parameters $Q_y, Q_u, Q_T, \underbar{$Q_r$}, \overline{Q_r}$ are 1, 1, 1, 100, 1 respectively. In this scenario, we put more weight on \underbar{$Q_r$} to make the system not violate the lower bound, which is more crucial for the zone control. As described in the previous sections, it is difficult to get back to normal again if the plant gets stressed once. The actual upper and lower bound of the system for the maximum yield and stress-free zone is -0.25 m and -3.1 m \cite{feddes_parameterizing_2004}. Things to note are that the stress-free zone depends upon different types of crop and ET values. To ensure the system does not experience the actual stress, a more conservative zone is considered with lower and upper zones -2.8 m and -1.0 m, respectively. So even if the system violates the more conservative zone, it does not experience the actual stress. The lower bound and upper bound of input are considered as 0 m/s and 4e-07 m/s, respectively. The lower bound of time is considered 30 min while the upper bound is kept as 16 days. 

\begin{figure}[!ht]%
\centering
\subfigure[]{
\includegraphics[width=0.46\columnwidth]{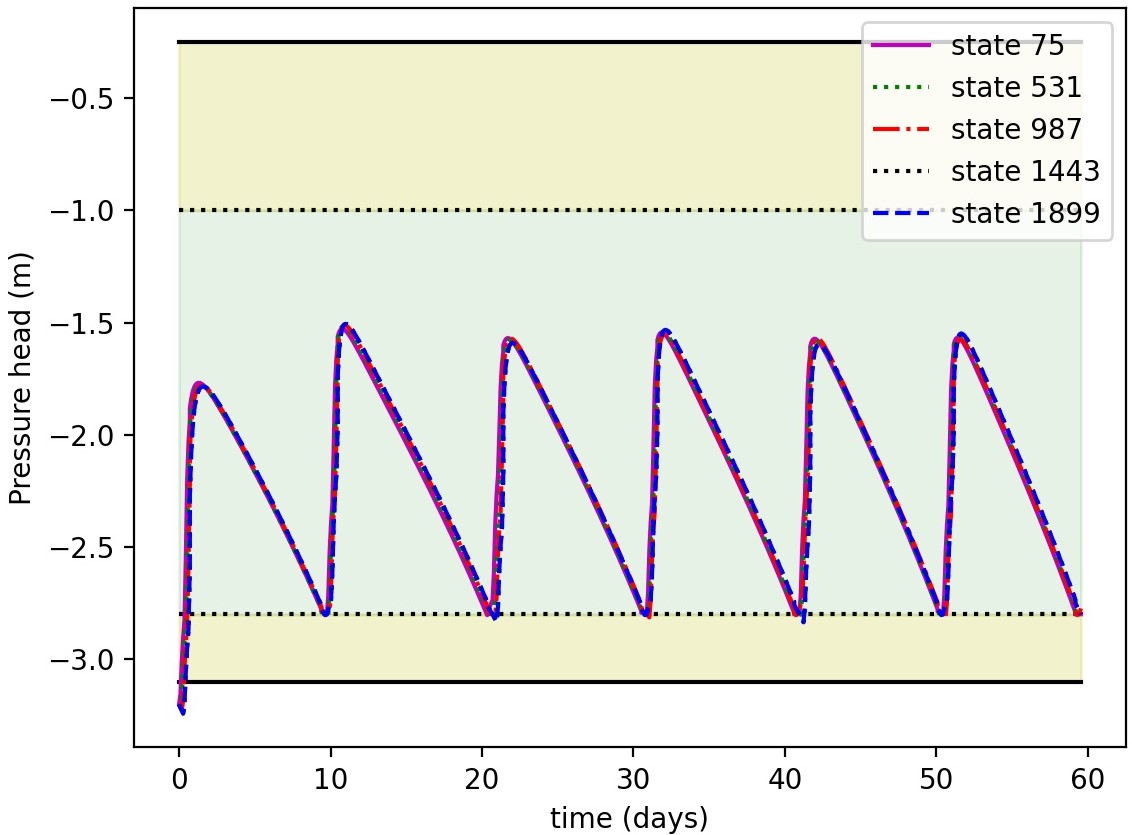}}
\qquad%
\subfigure[]{
\includegraphics[width=0.46\columnwidth]{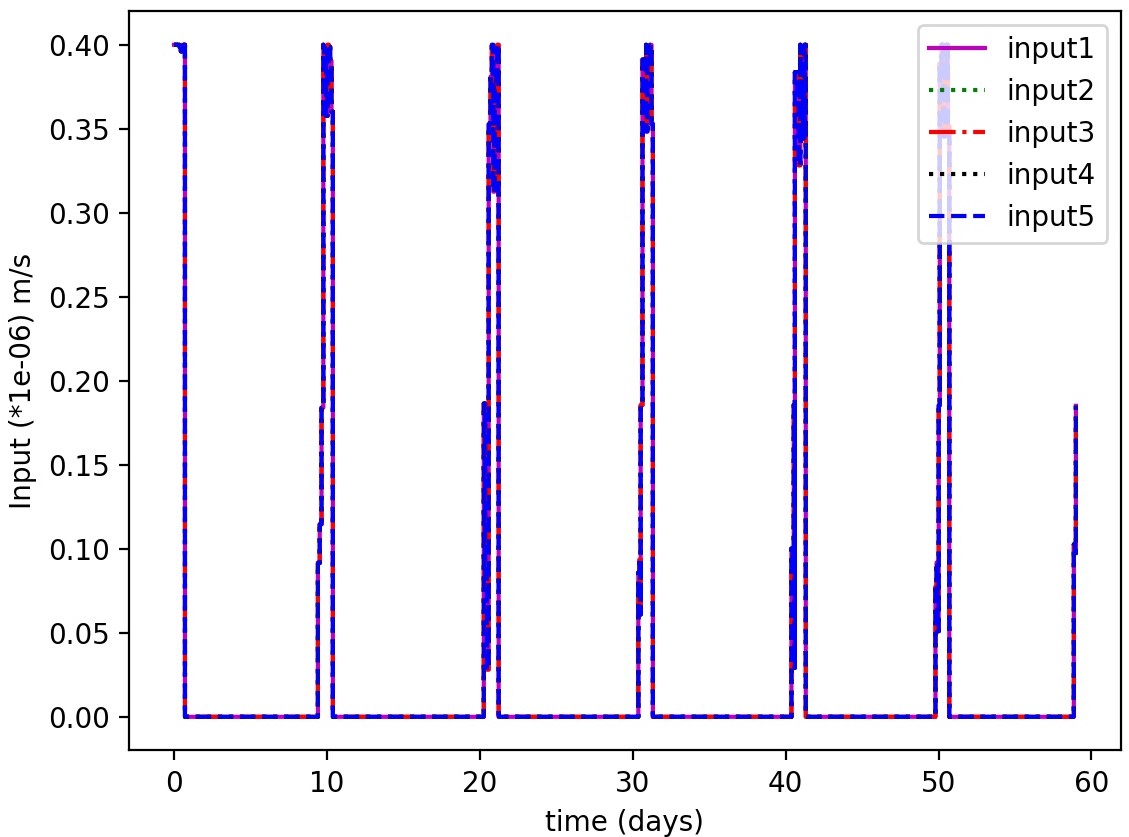}}
\caption{(a) Selected state trajectories under the proposed zone scheduler design for scenario 1; (b) Irrigation amount for 5 different sprinklers obtained from proposed zone scheduler}
\label{fig:case1_state}
\end{figure}

\begin{figure}[H]
\centerline{\includegraphics[width=0.5\columnwidth]{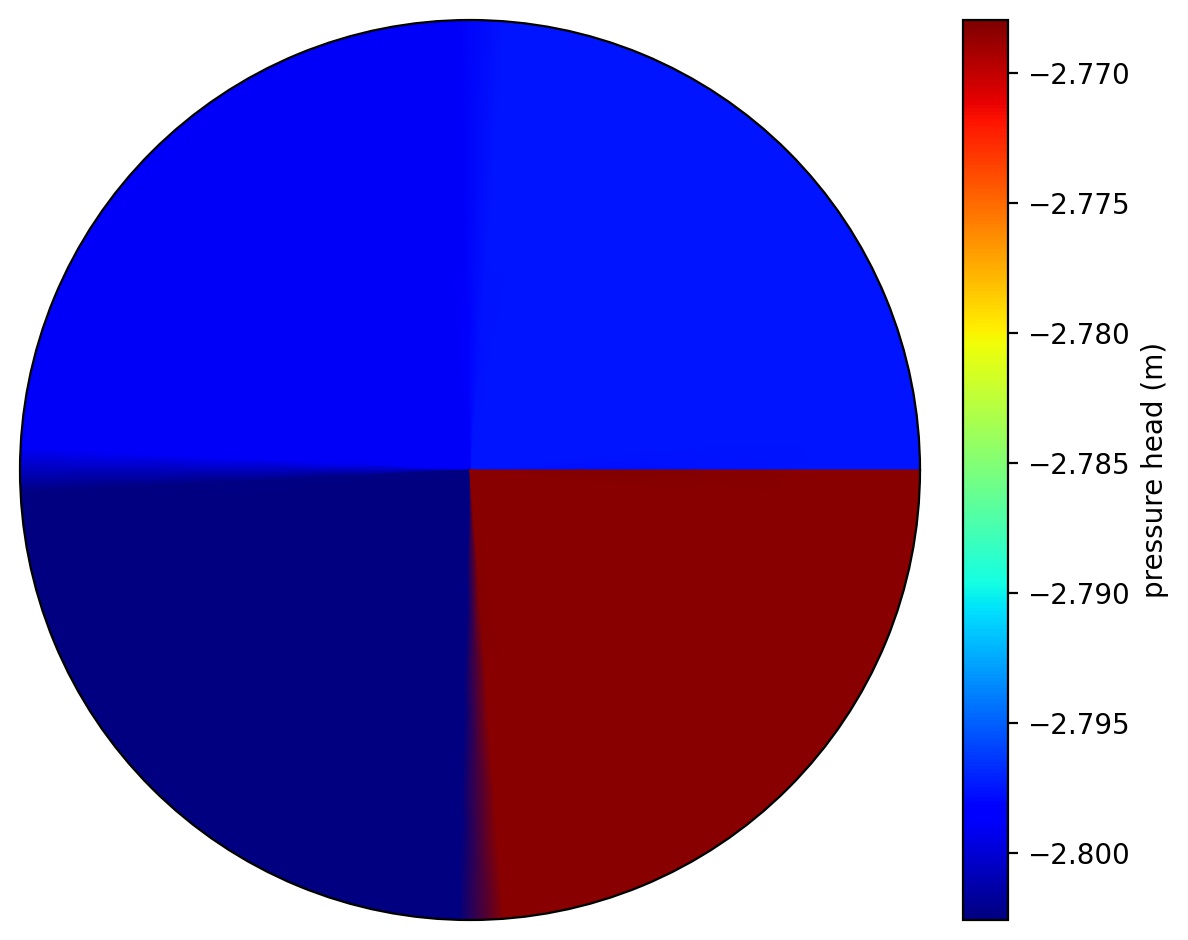}}
\caption{Pressure head values for root zone layer} 
\label{fig:case1_layer}
\end{figure}

Figure \ref{fig:case1_state}(a) shows the states trajectories of five randomly selected states on $2^{nd}$ layer and  Figure \ref{fig:case1_state}(b) shows the input amount. From Figure \ref{fig:case1_state}(a), we can observe that after it reaches the conservative lower zone, the system irrigates again. The root zone remains within the zone all the time so that the plants never get stressed. Figure \ref{fig:case1_state}(b) shows the input amount for all the five sprinklers. As the farm has a uniform soil type, all the sprinklers give nearly the same value. We can also observe that the irrigation event can be planned after nearly ten days interval. Figure \ref{fig:case1_layer} shows the $2^{nd}$ layer pressure head values. We can observe that all the nodes are around -2.8 m which is the conservative lower zone value. In all the quadrants, we see little bit different pressure head values because of the central pivot rotation. 

\begin{figure}[!ht]%
\centering
\subfigure[]{
\includegraphics[width=0.46\columnwidth]{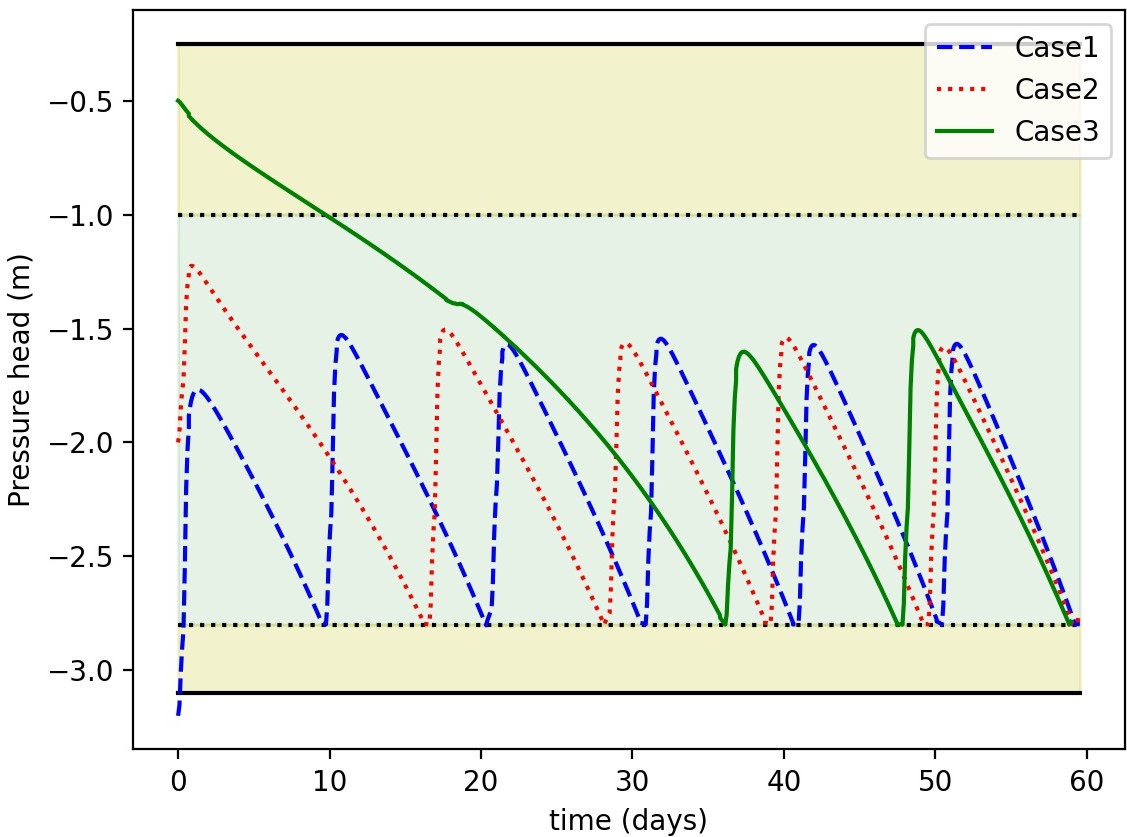}}
\qquad%
\subfigure[]{
\includegraphics[width=0.46\columnwidth]{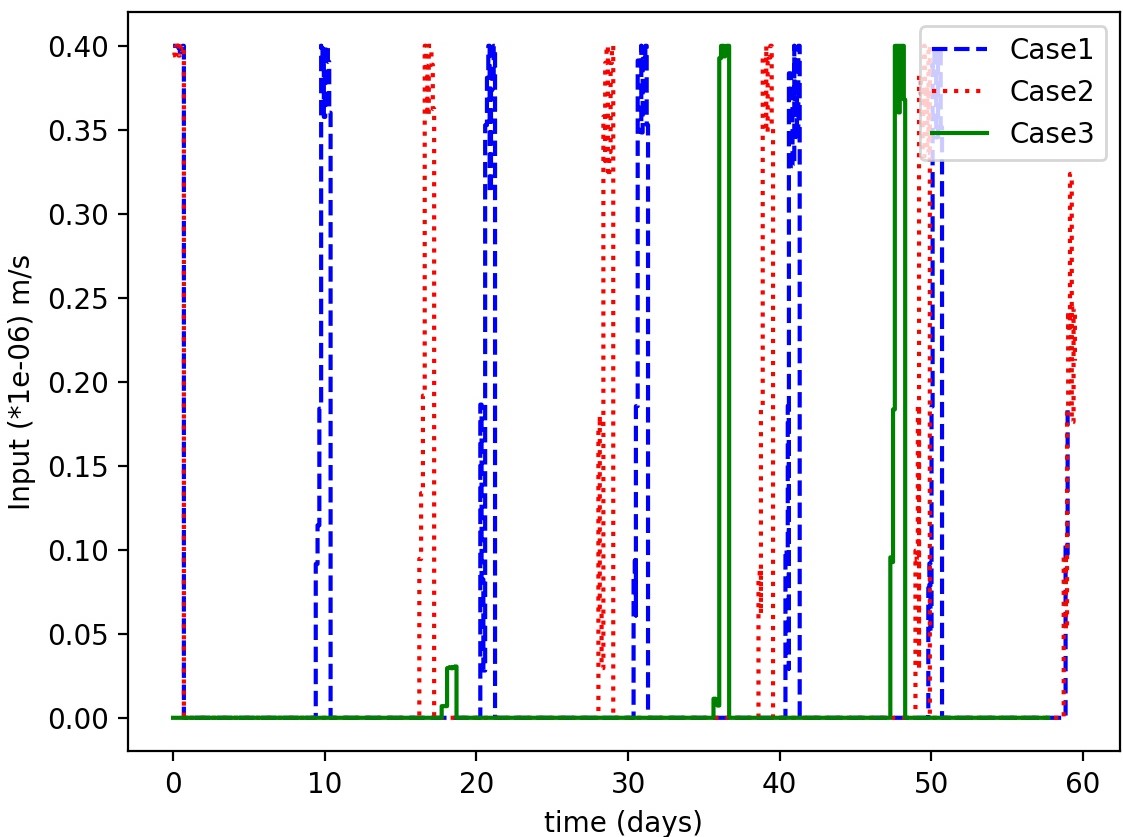}}
\caption{(a) Selected state trajectories for three cases (different initial conditions); (b) Input trajectories for three cases}
\label{fig:robust_case1}
\end{figure}

Further, the robustness of the proposed scheduler is analyzed. We consider three different cases based on initial conditions (-0.5 m, -3.2 m, and -2.0 m). The initial conditions are chosen such that two initial conditions are outside of the zone and one is already inside the zone. Figure \ref{fig:robust_case1}(a) presents the state trajectory of one randomly selected state 75 on the $2^{nd}$ layer. It can be observed that in all three cases, the scheduler works fine. Figure \ref{fig:robust_case1}(a) shows the input profile for sprinkler 1 for all the three cases. We can see that in case 1 the irrigation amount is higher than the other two cases. This is because the initial condition is outside the lower zone and requires more and frequent irrigation to keep within the zone. In case 3, as the initial condition is already very wet, the scheduler prescribes very little water until it reaches the lower zone. 

\subsubsection{Scenario 2}
In this scenario, we consider the non-uniform soil types, no rain, and constant ET. The motive is to show the efficiency of the scheduler in the presence of different soil types. The soil arrangement discussed in subsection \ref{sec:model_red} is considered in this scenario (Figure \ref{fig:para}). We consider $3^{rd}$ layer from the top (10 cm depth) as the root zone of the crop-type grass. The scheduler tries to keep only the $3^{rd}$ layer inside the zone. The values of the upper and lower bounds of the actual and conservative zones are the same as in scenario 1. The lower and upper bounds of inputs are 0 m/s and 4e-07 m/s. The lower and upper bounds of the time are 30 mins and 12 days.  

\begin{figure}[!ht]%
\centering
\subfigure[]{
\includegraphics[width=0.46\columnwidth]{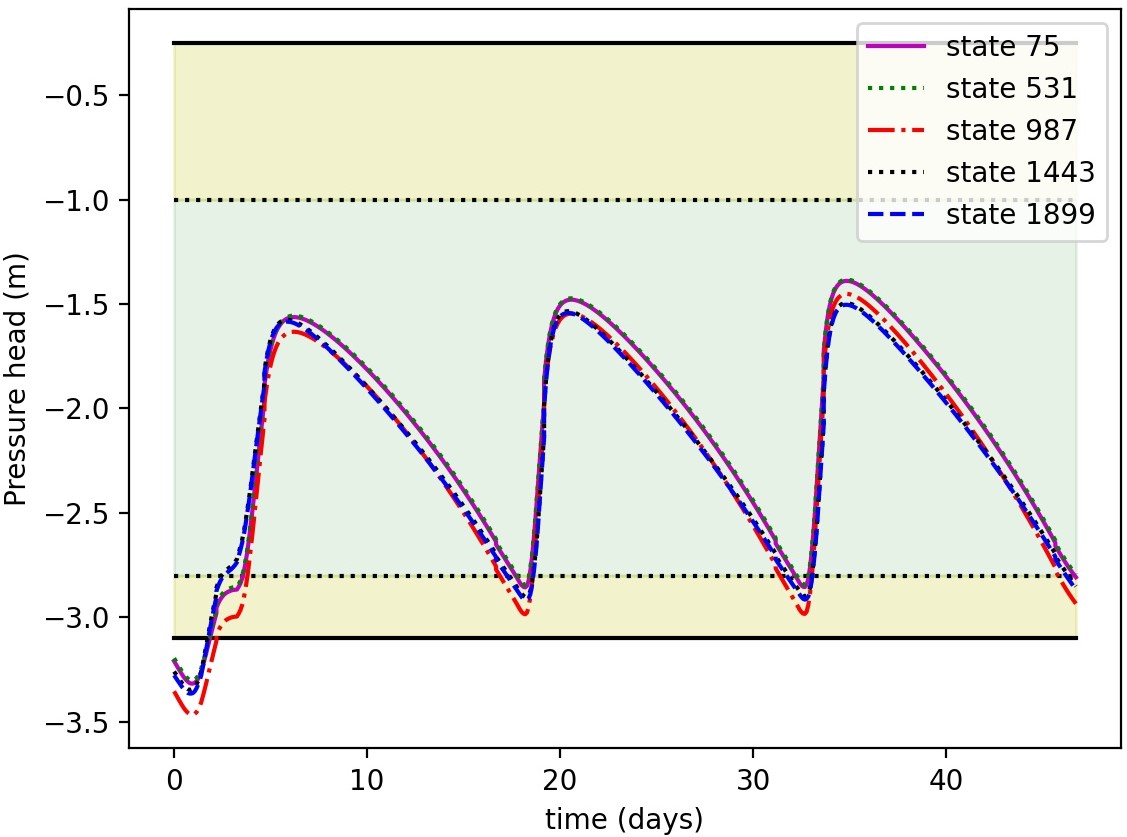}}
\qquad%
\subfigure[]{
\includegraphics[width=0.46\columnwidth]{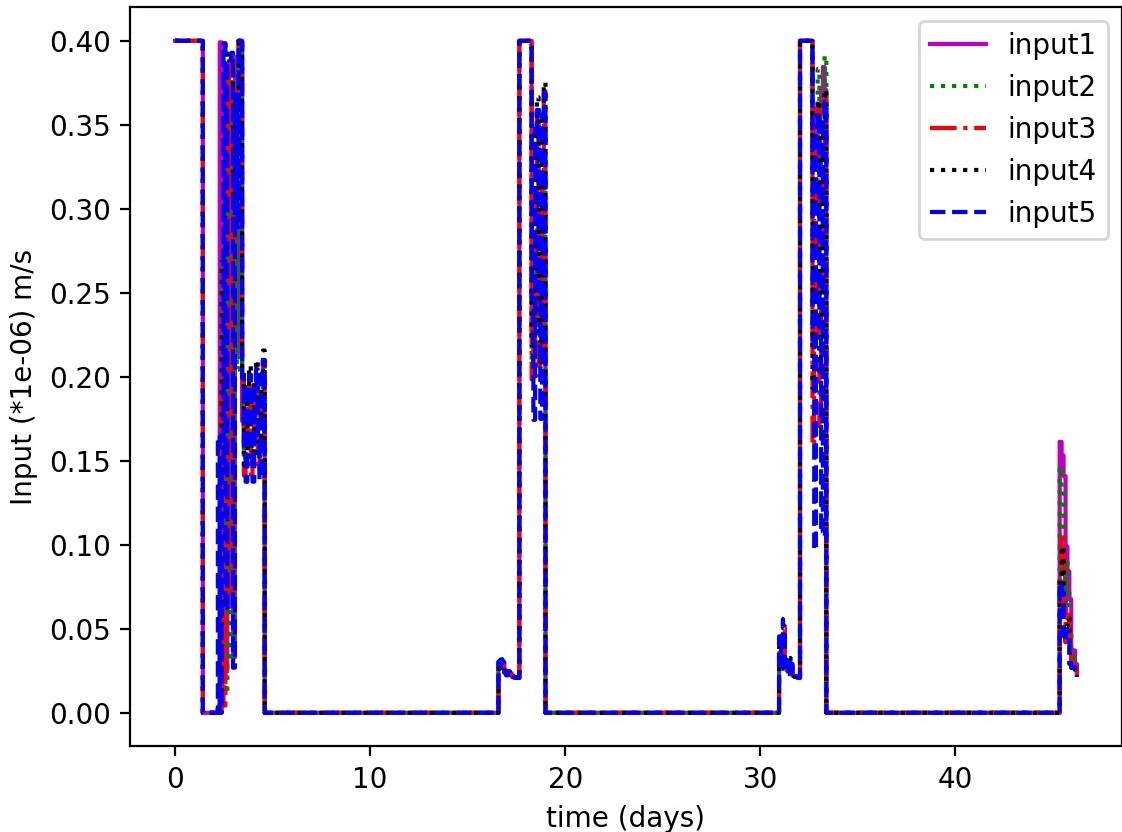}}
\caption{(a) Selected state trajectories at $3^{rd}$ layer under the proposed zone scheduler design for scenario 2; (b) Irrigation amount for 5 different sprinklers obtained from proposed zone scheduler}
\label{fig:case2:state}
\end{figure}

\begin{figure}[!ht]
\centerline{\includegraphics[width=0.5\columnwidth]{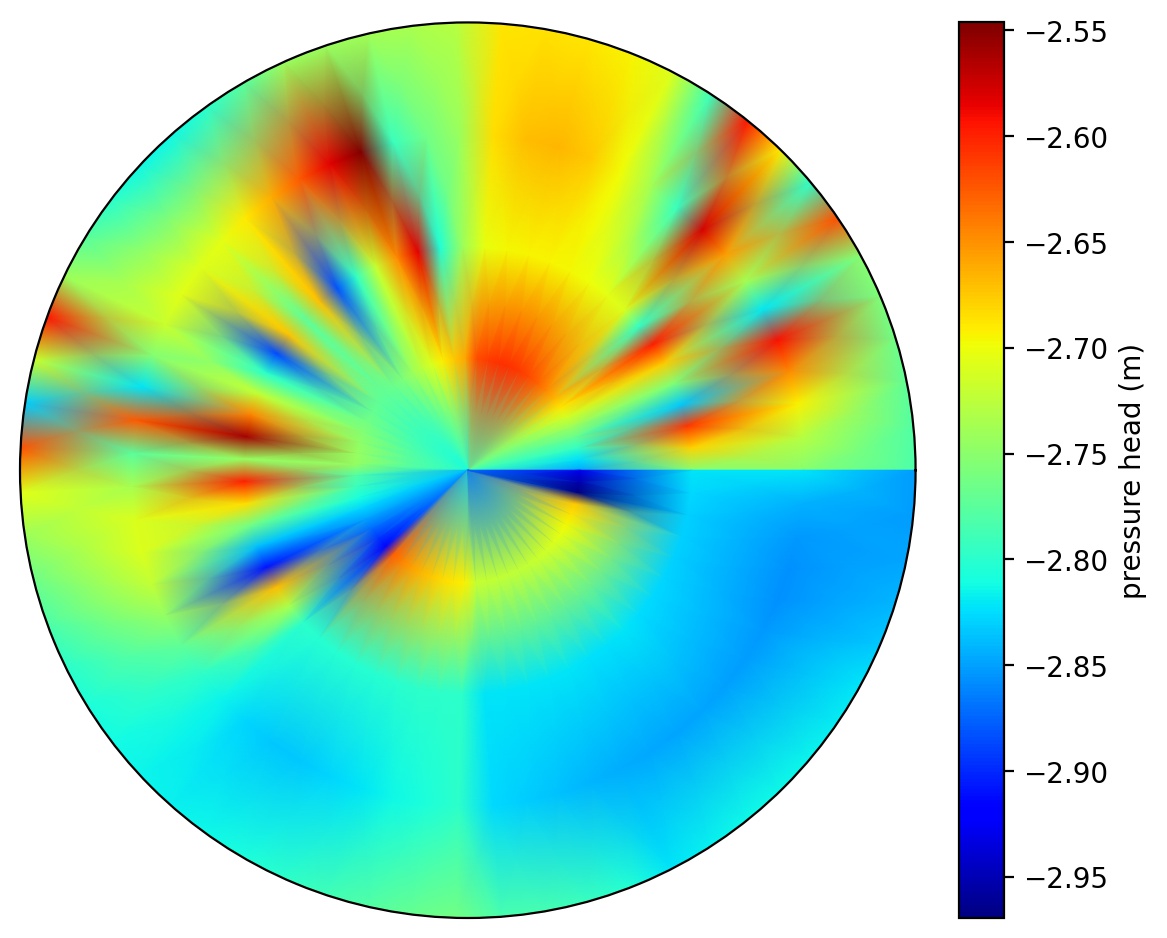}}
\caption{Pressure head values of depth 10cm} 
\label{fig:case2_layer}
\end{figure}

Figure \ref{fig:case2:state}(a) shows state trajectories of some selected states from the $3^{rd}$ layer. It shows that the states stay with in the zone and some of the states are little outside of the conservative zone. It may be caused because of the reduced model error and different soil types. All the states values for $3^{rd}$ layer at the end time is shown in Figure \ref{fig:case2_layer}. We can observe that the all the pressure head values are above the lower zone value -3.1 m. Figure \ref{fig:case2:state}(b) shows the input trajectories of five different sprinklers obtained form the scheduler. As observed from the figure, the input amounts are not same for all the sprinklers. This shows for different soil types, the sprinklers may put different amount of water.

\subsubsection{Scenario 3}
\begin{figure}[H]%
\centering
\subfigure[]{
\includegraphics[width=0.3\columnwidth]{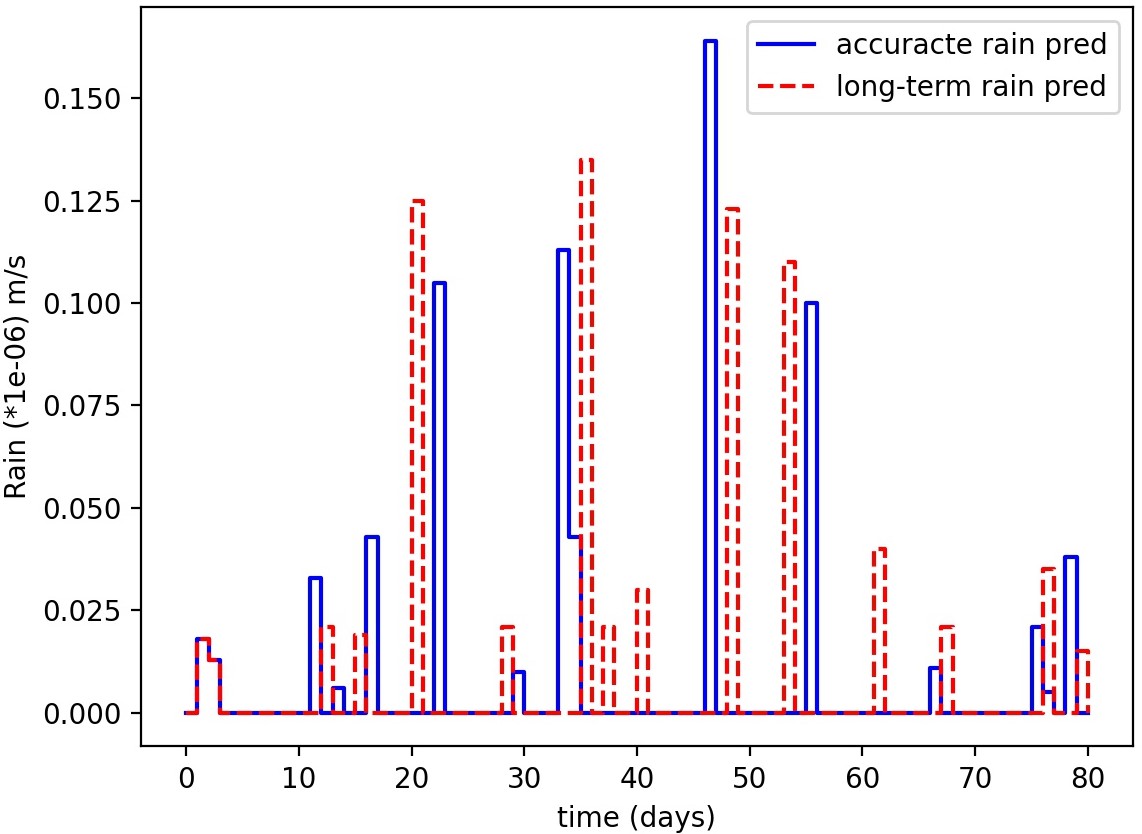}}
\qquad%
\subfigure[]{
\includegraphics[width=0.3\columnwidth]{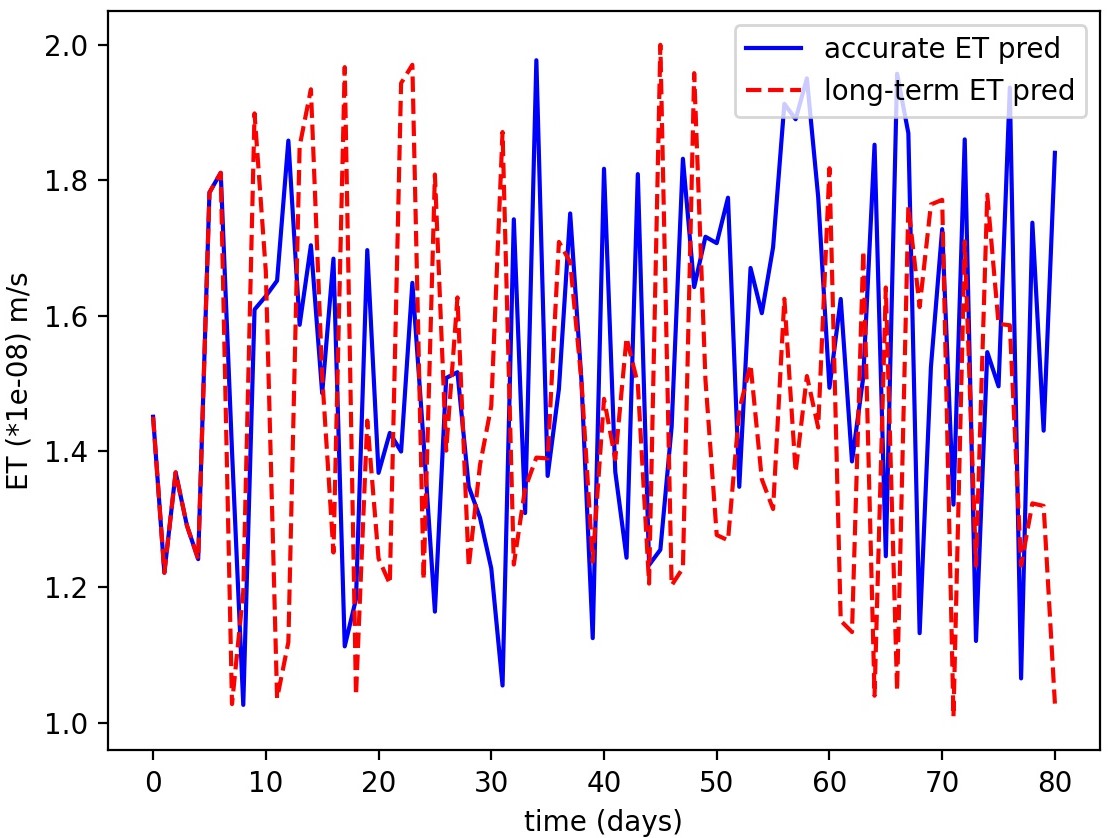}}
\subfigure[]{
\includegraphics[width=0.3\columnwidth]{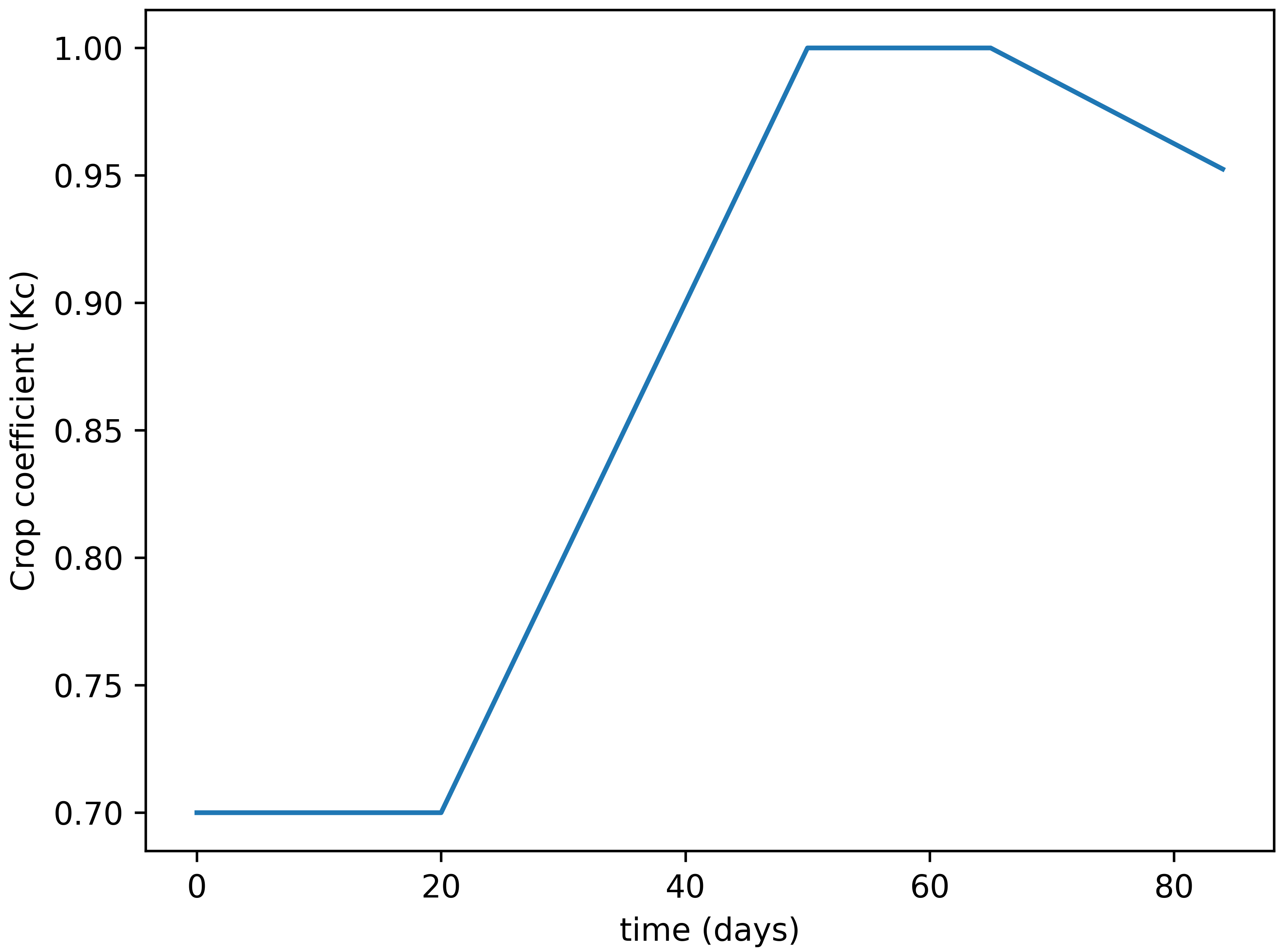}}

\caption{(a) Accurate rain prediction and long term rain prediction; (b) Accurate ET prediction and long-term ET prediction, (c) Crop coefficient for total growing season for lettuce}
\label{fig:weather}
\end{figure}

In this scenario, the variable ET and rain uncertainties are considered. The uniform soil type of loam and crop type of lettuce is chosen for the simulation. This scenario is shown to check the efficacy of the scheduler in the presence of weather disturbances and crop growth stages. As discussed in the crop modeling, 85\% root water is extracted from the top 30 cm soil. For the high yield and the crop not to experience any stress, keeping the top 30 cm in the stress-free zone is required. In this scenario, the objective of the scheduler is to keep all the layers in the zone. The values of the upper and lower bound for the actual zone are -0.25 m and -3.1 m. The upper and lower bound for the conservative zone is considered as -0.5 m and -2.3 m. The values of the tuning parameters $Q_y, Q_u, Q_T, \overline{Q_r}$, $\underbar{$Q_r$}$ are 1, 100, 1, 100, 0.01, 1 respectively. Things to note are that the tuning parameter values can be adjusted depending on the root growth with time. The lower and upper bounds of the time are 30 mins and 12 days. The upper and lower bound of the input are 0 m/s and 4e-07 m/s. In general, seven days rain predictions can be 80\% accurate. So for one horizon in the scheduler, the accurate weather prediction of 7 days is used, and for the rest of the days, the long-term prediction value is used. The values of accurate weather prediction and long-term weather prediction considered for this simulation are shown in Figure \ref{fig:weather}(a). Similarly, the reference ET value for accurate and long-term weather prediction is shown in Figure \ref{fig:weather}(b). The crop coefficient ($K_c$) for lettuce crop type for all the growing season is shown in \ref{fig:weather}(c). All growing season consists of the initial, crop development, mid-season, and late season. 

\begin{figure}[H]%
\centering
\subfigure[]{
\includegraphics[width=0.46\columnwidth]{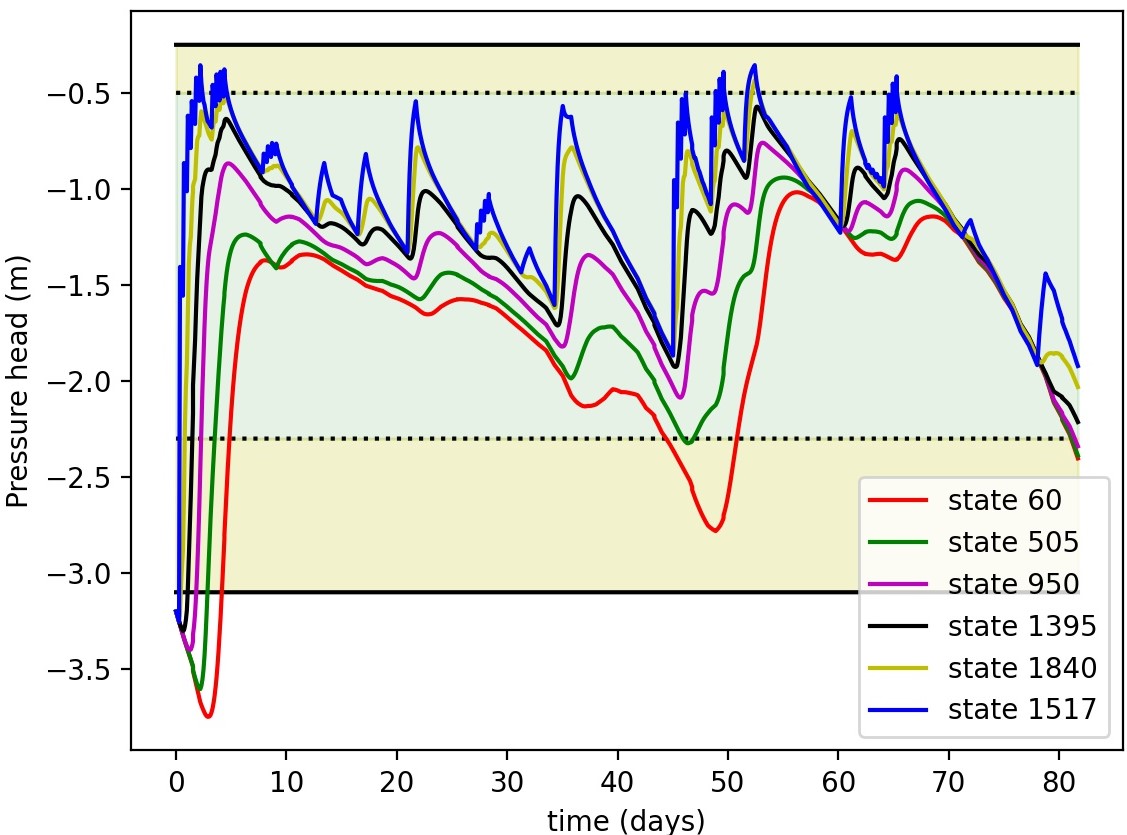}}
\qquad%
\subfigure[]{
\includegraphics[width=0.46\columnwidth]{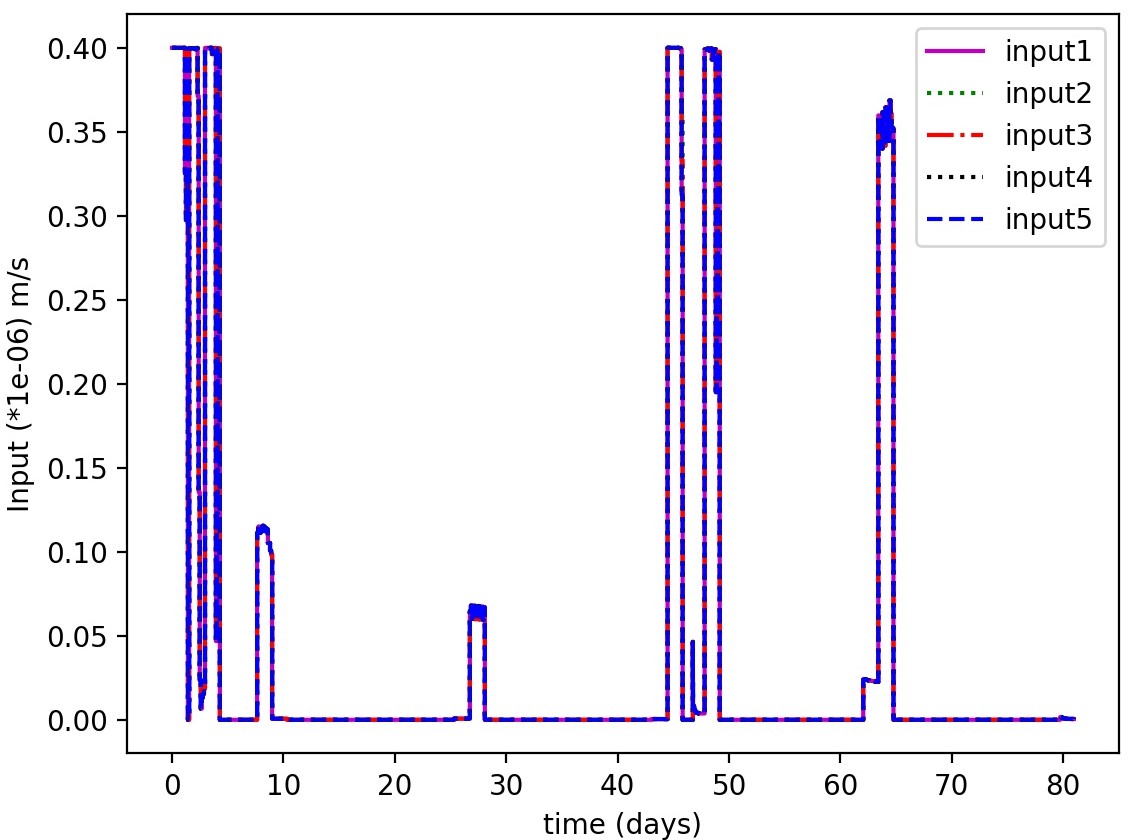}}
\caption{(a) Selected state trajectories for all layers under the proposed zone scheduler design for scenario 3; (b) Irrigation amount for 5 different sprinklers obtained from proposed scheduler for scenario 3}
\label{fig:sec3}
\end{figure}

Figure \ref{fig:sec3}(a) shows some selected state trajectories at different depths. We can observe that for all the depths, the states are in the stress-free zone. We can also see that around 50 days, state 60 (bottom node) goes outside of the conservative zone. This happens because around day 50, the crop is at the mature stage, and the ET values are high \ref{fig:weather}(c), and there is some delay between the water to reach the bottom node. That is why we choose a conservative zone of -2.3 m, So even if the bottom nodes go outside of the conservative zone, it may have less chance to go outside of the actual stress zone. \ref{fig:sec3}(b) shows the input amount for all five sprinklers. We can observe that there is frequent irrigation at the beginning because the states are outside of the zone. Moreover, around day 45-50, there is frequent irrigation because of high ET values, and the crop needs more water at that stage. For other days, because of the rain, the crops don't need much water.

\section{Conclusions}
In this article, the closed-loop scheduler for a large three-dimensional agro-hydrological system is proposed.  The cylindrical version of the finite difference model is explicitly constructed for an agricultural field equipped with a central pivot. The algorithm for the proposed structure-preserving model reduction is discussed. The motivation and the algorithm of the proposed scheduler are presented. The scheduler optimizes the irrigation amount and time of the following irrigation event to assure maximum yield, water preservation, and electricity usage. The proposed approach has been applied to three different cases. In all the cases, the proposed scheduler shows a satisfactory result. 
\section{Acknowledgment}
Financial support from Natural Sciences and Engineering Research Council of Canada and Alberta Innovates Technology Futures is gratefully acknowledged


\end{document}